\null

\input epsf
%*******************************************************************
%FONTS

\magnification 1200

%*******************************************************************
%FOOTNOTES - insert command \eightpoint to reduce

\newskip\ttglue

\font\eightrm=cmr8
\font\eighti=cmmi8
\font\eightsy=cmsy8
\font\eightbf=cmbx8
\font\eighttt=cmtt8
\font\eightsl=cmsl8
\font\eightit=cmti8
\font\sixrm=cmr6
\font\sixbf=cmbx6
\font\sixi=cmmi6
\font\sixsy=cmsy6

\def \eightpoint{\def\rm{\fam0\eightrm}% switch to 8-point type
\textfont0=\eightrm \scriptfont0=\sixrm \scriptscriptfont0=\fiverm
\textfont1=\eighti \scriptfont1=\sixi \scriptscriptfont1=\fivei
\textfont2=\eightsy \scriptfont2=\sixsy \scriptscriptfont2=\fivesy
\textfont3=\tenex \scriptfont3=\tenex \scriptscriptfont3=\tenex
\textfont\itfam=\eightit \def\it{\fam\itfam\eightit}%
\textfont\slfam=\eightsl \def\sl{\fam\slfam\eightsl}%
\textfont\ttfam=\eighttt \def\tt{\fam\ttfam\eighttt}%
\textfont\bffam=\eightbf \scriptfont\bffam=\sixbf
\scriptscriptfont\bffam=\fivebf \def\bf{\fam\bffam\eightbf}% \tt
\ttglue=.5em plus.25em minus.15em
\setbox\strutbox=\hbox{\vrule height7pt depth2pt width0pt}%
\normalbaselineskip=9pt
\let\sc=\sixrm \let\big=\eightbig \normalbaselines\rm }
%******************************************************************* 
%GREEK LETTERS
\def\a{\alpha}
\def\b{\beta}
\def\c{\gamma}
\def\d{\delta}
\def\e{\epsilon}

\def\l{\lambda}
\def\m{\mu}
\def\n{\nu}
\def\o{\theta}
\def\p{\pi}
\def\r{\rho}
\def\s{\sigma}

\def\C{\Gamma}
\def\D{\Delta}

%***************************************************************** 
%OTHER MACROS
\def\pl{\partial}
\def\ra{\rightarrow}

\def\DD{{\cal D}}

\def\OO{{\cal O}}

\def\GV{{\rm GeV}}
\def\MV{{\rm MeV}}

%***************************************************************** %TITLE PAGE

{\nopagenumbers

\line{\hfil SWAT 98/209 }
\line{\hfil CERN--TH/98--374 }
\vskip2cm
\centerline{\bf $U_A(1)$ PROBLEMS AND GLUON TOPOLOGY}
\vskip0.4cm
\centerline{\bf -- ANOMALOUS SYMMETRY IN QCD ${}^*$}
\vskip1.2cm
\centerline{\bf G.M. Shore}
\vskip0.7cm
\centerline{\it Department of Physics} 
\centerline{\it University of Wales Swansea} 
\centerline{\it Singleton Park}
\centerline{\it Swansea, SA2 8PP, U.K. }
\vskip0.3cm
\centerline{\it and} 
\vskip0.3cm
\centerline{\it Theory Division} 
\centerline{\it CERN}
\centerline{\it CH 1211 Geneva 23, Switzerland} 
\vskip1cm
\noindent{\bf Abstract}
\vskip0.3cm
Many of the distinctive and subtle features of the dynamics in the
$U_A(1)$ channel in QCD can be related to gluon topology, more precisely
to the topological susceptibility $\chi(k^2) 
= i\int d^4x~e^{ikx}\langle 0|T~Q(x)~Q(0)|0\rangle$,
where $Q = {\a_s\over8\pi} {\rm tr} G_{\m\n} \tilde G^{\m\n}$ is the gluon 
topological charge density. The link is the $U_A(1)$ axial (ABJ) anomaly.
In this lecture, we describe the anomalous $U_A(1)$ chiral Ward identities
in a functional formalism and show how two apparently unrelated 
`$U_A(1)$ problems' -- the mass of the $\eta'$ and the violation of
the Ellis-Jaffe sum rule in polarised deep-inelastic scattering -- can be
explained in terms of the gluon topological susceptibility. 
They are related through a $U_A(1)$ extension of the Goldberger-Treiman
formula, which is derived here for QCD with both massless and massive quarks.

\vskip1.5cm
\noindent ${}^*$ ~{\it Lecture presented at the 1998 Zuoz Summer School
on `Hidden Symmetries and Higgs Phenomena'.}

\vskip0.7cm

\line{SWAT 98/209 \hfil}
\line{CERN--TH/98--374 \hfil}
\vskip0.1cm
\line{November 1998 \hfil}

\vfill\eject } 

\pageno=1

\noindent {\bf 1. Introduction}
\vskip0.5cm
The central theme of this school is symmetry, in particular 
the realisation of `hidden' or `spontaneously broken' symmetries 
in quantum field theory.
Another especially interesting realisation of symmetry in QFT is
exemplified by the flavour singlet chiral $U_A(1)$ symmetry in 
QCD with massless quarks. This is a subtle case, because although
global chiral $U_A(1)$ is a symmetry of the classical lagrangian
of QCD, the corresponding current is not conserved in the quantum 
theory due to the well-known ABJ anomaly[1].
Nevertheless, this `anomalous symmetry' has important implications
for the phenomenology of QCD. 

In this lecture, we describe how to
analyse the consequences of the $U_A(1)$ symmetry in QCD, using the
formalism of anomalous chiral Ward identities. This formalism is then
used to relate two `$U_A(1)$ problems' -- the mass of the $\eta'$
and the violation of the Ellis-Jaffe sum rule[2] in polarised 
deep-inelastic scattering (DIS) -- to gluon topology in the form of 
the gluon topological susceptibility
$$ 
\chi(k^2) = i\int d^4x~e^{ikx}\langle 0|T~Q(x)~Q(0)|0\rangle
\eqno(1.1)
$$
where $Q = {\a_s\over8\pi} {\rm tr} G_{\m\n} \tilde G^{\m\n}$ is the gluon 
topological charge density. 

The topological susceptibility is a fundamental correlation function
in pure gluodynamics (Yang-Mills theory) or QCD itself and is the key
to understanding much of the distinctive dynamics in the $U_A(1)$ 
channel. It has been studied non-perturbatively using lattice gauge
theory[3-5], spectral sum rules[6-8], instanton models of the vacuum[9], 
etc., with the following results:
$$\eqalignno{
&{\rm Gluodynamics} ~~~~~~~~\chi^{YM}(0) \simeq -(180~\MV)^4 ~~~~~~~
\chi'_{YM}(0) \simeq -(10~\MV)^2  \cr
&{\rm QCD} ~(m_q=0) ~~~~~~~\chi(0) = 0 ~~~~~~~~~~~~~~~~~~~~~~~~~~~~
\chi'(0) \simeq (26~\MV)^2  
&(1.2) \cr }
$$
where $\chi'(0) = {d\over dk^2} \chi(k^2)\big|_{k=0}$.
In gluodynamics, the value $\chi^{YM}(0)$ is well-established[3,6]
and can be calculated in lattice gauge theory.
In QCD with massless quarks, the result $\chi(0) = 0$ is an exact identity,
following (as we show below) directly from the anomalous chiral Ward
identities.
The quoted value for $\chi'(0)$ in full QCD was obtained using
spectral sum rules[7,8]. Lattice techniques are still being refined[4,5]
to produce a reliable result.

The connection with topology arises as follows. $Q$ is a total divergence,
viz.
$$
Q = \pl^\m K_\m
\eqno(1.3)
$$
where $K_\m = {\a_s\over4\pi}\e_{\m\n\r\s}{\rm tr}\bigl(
A^\n G^{\r\s} - {1\over3}g A^\n[A^\r,A^\s]\bigr)$ 
is the Chern-Simons current. However, the integral
over (Euclidean) spacetime of $Q$ need not vanish. In fact, for gauge field
configurations which become pure gauge at infinity,
$$
\int d^4 x ~Q = n \in {\bf Z}
\eqno(1.4)
$$
where the integer $n$ is the topological winding number (technically, 
an element of the homotopy group $\pi_3(SU(N_c))$, where $SU(N_c)$ is
the gauge group[10, ch.~23]) or `instanton number'. Instantons
are classical field configurations which contribute to the path integral
for which eq.(1.4) gives $n\neq 0$.

The connection with the $U_A(1)$ current in QCD arises through the
famous ABJ anomaly. In a sense we shall make precise shortly, the
flavour singlet axial current $J_{\m5}^0$ is not conserved in the quantum
theory, even though it is the Noether current for the classical QCD
lagrangian with massless quarks, but satisfies the anomaly equation 
(for $m_q=0$)
$$
\pl^\m J_{\m5}^0 - 2n_f Q \sim 0
\eqno(1.5)
$$
where $n_f$ is the number of quark flavours. This identity provides
the link between the quark dynamics and gluon topology.
  
Our first example of a `$U_A(1)$ problem' concerns the mass of the
$\eta'$. This is much larger than would be the case if it were the
pseudo-Goldstone boson for spontaneously broken $U_A(1)$, as we
would expect in the absence of the anomaly. Indeed, for massless
QCD, the $\eta'$ would be an exact, massless Goldstone boson
but for the anomaly. In section 3, we explain how the Goldstone theorem
is circumvented by the anomaly and derive the Witten-Veneziano mass
formula[11,12] for the $\eta'$. This relates, to leading order in the
$1/N_c$ expansion, the mass of the $\eta'$ to the topological
susceptibility in pure gluodynamics.\footnote{${}^{(1)}$}{\eightpoint
\noindent In this lecture, we derive this and other results directly
from the anomalous chiral Ward identities. For the related approach
of using effective chiral lagrangians in the $1/N_c$ expansion to
study $\eta'$ physics, see e.g.~refs.[13,14,15].}
For QCD with massless quarks, the formula is simply
$$
m_{\eta'} \simeq {\sqrt6\over f_\pi} \sqrt{\chi^{YM}(0)}
\eqno(1.6)
$$

The second example concerns the much-publicised `proton spin' problem
(for a review, see e.g.~ref.[16]), i.e.~the violation of the Ellis-Jaffe
sum rule observed in measurements of the first moment of the polarised
proton structure function $g_1^p$ in deep inelastic scattering (DIS).
As explained in section 4, the first moment $\int dx g_1^p(x;Q^2)$
is related, through the OPE for the product of two electromagnetic
currents, to the proton matrix elements of the $SU(3)$ flavour singlet
and non-singlet axial currents $\langle p|J_{\m5}^a|p\rangle$,
for $a=0,3,8$. The corresponding `axial charges' of the proton are
denoted by $a^0(Q^2)$, $a^3$ and $a^8$, where the singlet charge (only)
has an explicit dependence on the scale $Q^2$ of the deep inelastic
process. This is due to the non-trivial renormalisation of the
singlet axial current induced by the anomaly. We have found[18,19] the 
following formula relating the ratio of the flavour singlet and 
non-singlet axial charges to the slope of the gluon topological 
susceptibility:
$$
{a^0(Q^2)\over a^8} \simeq {\sqrt6\over f_\pi} \sqrt{\chi'(0)}\big|_{Q^2}
\eqno(1.7)
$$
The similarity to eq.(1.6) is striking. This formula is the basis of our
quantitative resolution of the `proton spin' problem[7], which is explained
as a consequence of topological charge screening by the QCD vacuum.

Although apparently quite different, these two examples are in fact
intimately related through a $U_A(1)$ extension of the familiar 
Goldberger-Treiman relation. The essence of the conventional GT
relation is that it links the dynamics of the pseudovector and
pseudoscalar channels. This is just what we need to explain the
OZI violation at the heart of the `proton spin' problem, provided
an extension of the GT relation to the anomalous $U_A(1)$ can be
found. This was achieved in refs.[17,18,19] and underlies our result (1.7).
In section 5, we derive the $U_A(1)$ GT relation. For massless QCD,
we find
$$
2m_N G_A^0 = 2n_f \sqrt{\chi'(0)}~\hat \C_{\eta^0 NN}\big|_{k=0}
\eqno(1.8)
$$
(Here, $G_A^a$ is just an alternative notation for the nucleon axial
charges, with a more convenient $SU(3)$ normalisation.)
$\hat \C_{\eta^0 NN}$ is a vertex describing the coupling of the 
nucleon to the $\eta^0$ -- an unphysical state in QCD which in the
OZI limit (see section 3) is identified with the exact $U_A(1)$ Goldstone 
boson. For QCD with massive quarks and flavour $SU(3)$ breaking, the 
corresponding formula mixes the flavour singlet and non-singlet sectors.
In this case, we find[8]
$$
2m_N G_A^a ~=~  F_{ab} ~\hat \C_{\eta^b NN}\big|_{k=0}
\eqno(1.9) 
$$
where $F$ is determined from
$$
F_{ac} F_{cb}^T = \lim_{k=0} {d\over dk^2}~ i \int dx~e^{ikx}~\langle 0|
T~\pl^\m J_{\m5}^a(x)~\pl^\n J_{\n5}^b(0)|0\rangle
\eqno(1.10)
$$
providing a natural generalisation of eq.(1.8).

\vskip1cm

\noindent{\bf 2.~~Chiral Ward Identities and the Renormalisation Group}
\vskip0.5cm
To study the phenomenology of the $U_A(1)$ anomaly, we will need to find
the Green functions of the composite operators which couple to the
relevant physical states. These are the currents and pseudoscalar 
operators $J_{\m5}^a$, $Q$, $\phi_5^a$ together with $\phi^a$ where
$$\eqalignno{
J_{\m5}^a &= \bar q \c_\m \c_5 T^a q ~~~~~~~~~~~~~~~~~
Q = {\a_s\over8\pi} {\rm tr} G_{\m\n} \tilde G^{\m\n} \cr 
\phi_5^a &= \bar q \c_5 T^a q ~~~~~~~~~~~~~~~~~~~~
\phi^a = \bar q T^a q
&(2.1) \cr }
$$
$G_{\m\n}$ is the field strength for the gluon field $A_\m$.
In this notation, $T^i = {1\over2}\l^i$ are flavour $SU(n_f)$ generators,
and we include the singlet $U_A(1)$ generator $T^0 = {\bf 1}$ and let the
index $a = 0, i$.
We will only need to consider fields where $i$ corresponds to a generator
in the Cartan sub-algebra, so that $a = 0, 3, 8$ for $n_f = 3$ quark
flavours. We define $d$-symbols by $\{T^a,T^b\} = d_{abc} T^c$. Since this
includes the flavour singlet $U_A(1)$ generator, they are only symmetric on
the first two indices. For $n_f = 3$, the explicit values are $d_{000} =
d_{033} = d_{088} = 2, d_{330} = d_{880} = 1/3, d_{338} = d_{383} =
-d_{888} = 1/\sqrt3$. (For further notation and description of the
formalism used here, see refs.[19,8].)

The Green functions, or correlation functions, are constructed from the
generating functional $W[V_{\m5}^a, \o, S_5^a, S^a]$, where
$V_{\m5}^a, \o, S_5^a, S^a$ are the sources for the composite
operators $J_{\m5}^a, Q, \phi_5^a, \phi^a$ respectively.
Functional derivatives of $W$ yield Green functions which are `1PI'
w.r.t.~the designated fields (composite operators). Explicitly,
$$
e^{iW} ~=~ \int \DD A \DD\bar q \DD q ~\exp i\int d^4x\bigl(
{\cal L}_{\rm QCD} + V_{\m5}^a J_{\m5}^a + \o Q + S_5^a \phi_5^a
+ S^a \phi^a\bigr)
\eqno(2.2)
$$

The chiral Ward identities are found by exploiting the invariance
of $W$ under a change of variables in the path integral corresponding
to a chiral transformation $q \rightarrow e^{i\a T^a\c_5}q$. 
This gives
$$
\int \DD A \DD\bar q \DD q~\Bigl[\pl^\m J_{\m5}^a - 2n_f \d^{a0} Q
-d_{adc}m^d\phi_5^c - \d \Bigl(i\int d^4x {\cal L}_{\rm QCD}\Bigr)\Bigr]~
\exp i\int d^4x \bigl( \ldots \bigr) ~~=~~0
\eqno(2.3)
$$
The terms in the square bracket are simply those arising from
Noether's theorem, which defines the symmetry current through a 
functional derivative of the action, with the addition of the anomaly term.
This can be understood as arising from the non-invariance of the 
path integral measure $\DD\bar q \DD q$ in a background gauge field $A_\m$.
A careful derivation of the anomaly by this method can be found in
several standard textbooks on QFT, e.g.~ref.[10, ch.~22].
The chiral variation term w.r.t.~the elementary fields is then simply
re-expressed as a variation w.r.t.~the sources, giving finally the
functional form of the (anomalous) chiral Ward identities:
$$
\pl_\m {\d W\over\d V_{\m5}^a} - 2n_f \d_{a0}{\d W\over\d\o}-d_{adc} m^d
{\d W\over \d S_5^c}
+ d_{adc} S^d {\d W \over \d S_5^c}
- d_{adc} S_5^d {\d W\over\d S^c}
= 0
\eqno(2.4)
$$
This is the key equation which will be the basis of all the results 
derived in this lecture. It makes precise the anomaly equation (1.5).

It will be useful in what follows to introduce some streamlined notation.
The quark mass matrix is written as $m^a T^a$, so that for $n_f=3$, 
$$
\left(\matrix{m_u &0 &0 \cr
0 &m_d &0 \cr
0 &0 &m_s \cr}\right)
= m^0 {\bf 1} + m^3 T^3 + m^8 T^8
\eqno(2.5)
$$
The chiral symmetry breaking condensates may be similarly written as 
$$
\left(\matrix{ \langle \bar u u\rangle &0 &0 \cr 0 &\langle \bar d d\rangle
&0 \cr
0 &0 &\langle \bar s s\rangle \cr}\right) = {1\over3} \phi^0 {\bf 1} + 2
\phi^3 T^3 + 2 \phi^8 T^8 
\eqno(2.6)
$$
where $\langle \phi^c\rangle$ is the VEV $\langle \bar q T^c q\rangle$.
It is also convenient to introduce the still more compact notation 
$$
M_{ab} = d_{acb} m^c
~~~~~~~~~~~~~~~~~~
\Phi_{ab} = d_{abc} \langle \phi^c\rangle 
\eqno(2.7)
$$
for which
$$
{1\over8} {\rm det} M = m_u m_d m_s
~~~~~~~~~~~~~
{1\over6} {\rm det} \Phi =
\langle \bar u u \rangle ~\langle \bar d d \rangle ~\langle \bar s s
\rangle 
\eqno(2.8)
$$

Notice that the derivation of the chiral Ward identities sketched above
was entirely in terms of the bare operators. Renormalised composite 
operators are defined as follows[20]:
$$\eqalignno{
&J_{\m5R}^0 = Z J_{\m5B}^0  ~~~~~~~~~~
J_{\m5R}^{a\neq0} =  J_{\m5B}^{a\neq0} \cr
&Q_R = Q_B - {1\over 2n_f}(1-Z) \pl^\m J_{\m5B}^0 \cr
&\phi_{5R}^a = Z_\phi \phi_{5B}^a ~~~~~~~~~~~~~
\phi_R^a = Z_\phi \phi_B^a 
&(2.9) \cr}
$$
where $Z_\phi$ is the inverse of the mass renormalisation, 
$Z_\phi = Z_m^{-1}$. The anomalous dimensions associated with $Z$ and 
$Z_\phi$ are denoted $\c$ and $\c_\phi$ respectively. 
Notice the mixing of the operator $Q$ with $\pl^\m J_{\m5}^0$ under
renormalisation. Most importantly, with these definitions the combination 
$\pl^\m J_{\m5}^0 - 2n_f Q$ occurring in the $U_A(1)$ anomaly equation
is RG invariant. The chiral Ward identities therefore take precisely
the same form expressed in terms of the bare or renormalised operators.
From now on, therefore, we use eq.(2.4) as an identity for renormalised
composite operators (omitting the label `R' for notational simplicity).

It is also convenient to use a condensed notation where a functional
derivative is represented simply by a subscript, with a spacetime integral
assumed where appropriate. Also transforming to momentum space, we 
therefore write eq.(2.4) compactly as
$$
ik_\m W_{V_{\m5}^a} - 2n_f \d_{a0} W_{\o} - M_{ac} W_{S_5^c} + d_{adc}
S^d W_{S_5^c} - d_{adc} S_5^d W_{S^c}
= 0
\eqno(2.10)
$$
The Ward identities for composite operator Green functions are derived by
taking functional derivatives of this basic identity. We will need the
following identities for 2-point functions:
$$\eqalignno{
&ik_\m W_{V_{\m5}^a V_{\n5}^b} - 2n_f \d_{a0} W_{\o V_{\n5}^b} - M_{ac}
W_{S_5^c V_{\n5}^b} = 0 \cr
&ik_\m W_{V_{\m5}^a \o} - 2n_f \d_{a0} W_{\o \o} - M_{ac} W_{S_5^c \o} = 0 \cr
&ik_\m W_{V_{\m5}^a S_5^b} - 2n_f \d_{a0} W_{\o S_5^b} - M_{ac} W_{S_5^c
S_5^b} - \Phi_{ab} = 0 
&(2.11) \cr }
$$
 Combining the individual equations in (2.11), we find the important identity: 
$$
k_\m k_\n W_{V_{\m5}^a V_{\n5}^b} - M_{ac} \Phi_{cb} = W_{S_D^a S_D^b}
\eqno(2.12)
$$
where $S_D^a$ is the source for the current divergence operator 
$D^a = 2n_f \d_{a0} Q + M_{ac}\phi_5^c$. In canonical notation,
$$\eqalignno{
W_{S_D^a S_D^b} &= i \int dx~e^{ikx}~\langle 0| T~D^a(x)~D^b(0)|0\rangle
\cr &= i \int dx~e^{ikx}~\langle 0| T~\pl^\m J_{\m5}^a(x)~\pl^\n
J_{\n5}^b(0)|0\rangle 
&(2.13) \cr }
$$

The zero-momentum Ward identities play a special role. These follow
immediately from eqs.(2.11) under the assumption that there are no massless
particles (in particular, no exact Goldstone bosons) contributing $1/k^2$
poles in the 2-point functions. With this assumption, we find simply
$$\eqalignno{
&2n_f \d_{a0} W_{\o\o} + M_{ac} W_{S_5^c \o} = 0 \cr 
&2n_f \d_{a0} W_{\o S_5^b} + M_{ac} W_{S_5^c S_5^b} + \Phi_{ab} = 0 
&(2.14) \cr }
$$
Combining these, we find that the topological susceptibility
$\chi(0) \equiv W_{\o\o}(0)$ satisfies the identity
$$
(2n_f)^2 \chi(0)  =  M_{0b}M_{0c}W_{S_5^b S_5^c} + M_{0b} \Phi_{0b}
\eqno(2.15)
$$

\vskip0.3cm
Another key ingredient in the discussion of the `proton spin' problem and 
GT relations in sections 4 and 5 is the use of proper vertices for this
set of operators. These are defined as functional derivatives of a
generating functional $\C$, which is itself constructed from $W$ by
a partial Legendre transform in which the transform is made only on 
the fields $Q, \phi_5^a, \phi^a$ and not on the currents. The resulting 
proper vertices are 1PI w.r.t.~the propagators for these composite 
operators only. As explained fully in refs.[19,21], by separating off the 
particle poles in the propagators, this is the definition which gives 
the closest identification of these field-theoretic vertices with 
physical low-energy couplings such as e.g. $g_{\pi NN}$. 
We therefore define the generating functional $\C[V_{\m5}^a, Q,
\phi_5^a, \phi^a]$ as:
$$
\C[V_{\m5}^a, Q, \phi_5^a, \phi^a] = W[V_{\m5}^a, \o,
S_5^a, S^a] - \int dx~\Bigl(\o Q + S_5^a \phi_5^a + S^a \phi^a \Bigr) 
\eqno(2.16)
$$

The chiral Ward identities corresponding to eq.(2.10) are therefore: 
$$
ik_\m \C_{V_{\m5}^a} - 2n_f \d_{a0} Q - M_{ac} \phi_5^c + d_{acd}
\phi^d \C_{\phi_5^c} - d_{acd} \phi_5^d \C_{\phi^c} = 0 
\eqno(2.17)
$$
The Ward identities for the 2-point vertices will also be important. 
These follow directly from eq.(2.17):
$$\eqalignno{
&ik_\m \C_{V_{\m5}^a V_{\n5}^b} + \Phi_{ac} \C_{\phi_5^c V_{\n5}^b} = 0 \cr
&ik_\m \C_{V_{\m5}^a Q} - 2n_f \d_{a0} + \Phi_{ac} \C_{\phi_5^c Q} = 0 \cr
&ik_\m \C_{V_{\m5}^a \phi_5^b} + \Phi_{ac} \C_{\phi_5^c \phi_5^b} - M_{ab}
= 0 
&(2.18) \cr }
$$
It is then straightforward to derive the following important identity,
analogous to eq.(2.12):
$$
k_\m k_\n \C_{V_{\m5}^a V_{\n5}^b} + M_{ac} \Phi_{cb} = \Phi_{ac}
\C_{\phi_5^c \phi_5^d} \Phi_{db} 
\eqno(2.19)
$$

\vskip0.3cm
The renormalisation group equations for these quantities also play
a key role in understanding the physics of the $U_A(1)$ channel.
We therefore include here a brief and somewhat novel discussion
of the RGEs for the Green functions and proper vertices of these
composite operators in a functional formalism. For further details,
see especially refs.[22,19].

The fundamental RGE for the generating functional $W$ follows immediately 
from the definitions (2.9) of the renormalised composite operators.
It is:
$$
\DD W = \c\Bigl(V_{\m5}^0 - {1\over2n_f}\pl_\m \theta\Bigr)W_{V_{\m5}^0}
+ \c_\phi\Bigl(S_5^a W_{S_5^a} + S^a W_{S^a}\Bigr) + \ldots
\eqno(2.20)
$$
where $\DD = \Bigl(\m{\pl\over\pl\m} + \b{\pl\over\pl g} - \c_m\sum_q
m_q{\pl\over\pl m_q}\Bigr)\Big|_{V,\theta,S_5,S}$.
The notation $+\ldots$ refers to the additional terms which are required 
to produce the contact term contributions to the RGEs for $n$-point
Green functions of composite operators. These are discussed fully
in refs.[22,19], but will be omitted here for simplicity. They vanish at 
zero-momentum.

The RGEs for Green functions are found simply by differentiating
eq.(2.20) w.r.t.~the sources. Simplifying the results using the
chiral Ward identities (2.11), we find a complete set of RGEs for the
2-point functions. These are:
$$\eqalignno{
&\DD W_{V_{\m5}^0 V_{\n5}^0} = 2\c W_{V_{\m5}^0 V_{\n5}^0} +\ldots ~~~~~
\DD W_{V_{\m5}^0 V_{\n5}^b} = \c W_{V_{\m5}^0 V_{\n5}^b} +\ldots~~~~~
\DD W_{V_{\m5}^a V_{\n5}^b} = 0 +\ldots \cr
&\DD W_{V_{\m5}^0 \theta} = 2\c W_{V_{\m5}^0 \theta} 
+ \c {1\over2n_f} M_{0b} W_{V_{\mu5}^0 S_5^b} +\ldots \cr
&\DD W_{V_{\m5}^a \theta} = \c W_{V_{\m5}^a \theta}
+ \c {1\over2n_f} M_{0b} W_{V_{\mu5}^0 S_5^b} +\ldots    \cr
&\DD W_{V_{\mu5}^0 S_5^b} = (\c + \c_\phi) 
W_{V_{\mu5}^0 S_5^b} +\ldots ~~~~~~~~~
\DD W_{V_{\mu5}^a S_5^b} = \c_\phi W_{V_{\mu5}^a S_5^b} +\ldots \cr
&\DD W_{\theta \theta} = 2\c W_{\theta \theta} 
+ 2\c {1\over2n_f} M_{0b} W_{\theta S_5^b} +\ldots \cr
&\DD W_{\theta S_5^b} = (\c + \c_\phi) W_{\theta S_5^b} 
+ \c {1\over2n_f} \bigl(M_{0c} W_{S_5^c S_5^b}  + \Phi_{0b}\bigr) 
+\ldots \cr
&\DD W_{S_5^a S_5^b} = 2\c_\phi W_{S_5^a S_5^b} +\ldots
&(2.21) \cr}
$$
It is straightforward to check the self-consistency of these RGEs 
with the Ward identities (2.11) and (2.14). The pattern of cancellations 
which ensures this is nevertheless quite intricate.

Next, we need the RGE for the generating functional of the 1PI vertices.
This follows immediately from its definition in eq.(2.16) and the 
RGE (2.20) for $W$:
$$
\tilde\DD\C = \c\Bigl(V_{\m5}^0 - {1\over2n_f}\C_Q\pl_\m\Bigr)
\C_{V_{\m5}^0} - \c_\phi\Bigl(\phi_5^a \C_{\phi_5^a} + 
\phi^a \C_{\phi^a}\Bigr) +\ldots
\eqno(2.22)
$$
where $\tilde\DD = \Bigl(\m{\pl\over\pl\m} + \b{\pl\over\pl g} 
- \c_m\sum_q m_q{\pl\over\pl m_q}\Bigr)\Big|_{V,Q,\phi_5,\phi}$.

\vfill\eject
The RGEs for the 1PI vertices are found by differentiation, and using
the Ward identities (2.18) to simplify the results, we find
for the pseudoscalar sector:
$$\eqalignno{
\DD\C_{QQ} &= -2\c \C_{QQ} + 
2\c {1\over2n_f} \Bigl[\Phi_{0c} \C_{QQ} \C_{\phi_5^c Q}\Bigr] +\ldots \cr
\DD\C_{Q\phi_5^b} &= -(\c +\c_\phi) \C_{Q\phi_5^b} +
\c {1\over2n_f}\Bigl[\Phi_{0c}\bigl(
\C_{QQ}\C_{\phi_5^c \phi_5^b} + \C_{Q \phi_5^c} \C_{Q\phi_5^b}\bigr)
- M_{0b}\C_{QQ} \Bigr] +\ldots \cr
\DD\C_{\phi_5^a\phi_5^b} &= -2\c_\phi \C_{\phi_5^a\phi_5^b}
+ \c {1\over2n_f} \Bigl[\Phi_{0c}\C_{\phi_5^a Q}\C_{\phi_5^c\phi_5^b} 
- M_{0b}\C_{\phi_5^a Q} + ~a\leftrightarrow b~\Bigr]
+ \ldots
&(2.23) \cr}
$$
Here, $\DD = \tilde\DD + \c_\phi \langle\phi^a\rangle{\d\over\d\phi^a}$.
As explained in ref.[19], this is identical to the RG operator
$\DD$ defined above (acting on $W$) when the sources are set to 
zero and the fields to their VEVs.

It will also be useful to know the RGEs for the 3-point vertices
coupling a pseudoscalar operator to the nucleon. In the same way,
we find[19,8]
$$
\eqalignno{
&\DD \hat\C_{QNN} = - \c \hat\C_{QNN}
+ \c {1\over2n_f} \Bigl[\Phi_{0b} \Bigl(\C_{Q \phi_5^b} \hat\C_{QNN}
+ \C_{QQ} \hat\C_{\phi_5^b NN} \Bigr)\Bigr] +\ldots \cr
&\DD \hat\C_{\phi_5^a NN} = - \c_\phi \hat\C_{\phi_5^a NN}
+ \c {1\over2n_f} \Bigl[\Phi_{0b} \Bigl(\C_{\phi_5^a \phi_5^b} 
\hat\C_{QNN} + \C_{\phi_5^a Q} \hat\C_{\phi_5^b NN} \Bigr) 
- M_{0a} \hat\C_{QNN}\Bigr] + \ldots \cr
&{}&(2.24) \cr }
$$

These RGEs play two roles in the discussion that follows.
First, they will be used as consistency checks on the various
formulae we derive. Second, and most important, they will provide
the clue to identifying quantities which are likely to show violations
of the OZI rule and those for which we may reasonably expect the
OZI limit to be a good approximation. This is because we can identify
quantities which will be particularly sensitive to the $U_A(1)$
anomaly as those which have RGEs involving the anomalous dimension $\c$.

\vskip1cm

\noindent{\bf 3.~~Pseudoscalar Mesons and the Witten-Veneziano Mass
Formula for $\eta'$}
\vskip0.5cm
The anomalous chiral Ward identities have some immediate and important
consequences. For simplicity, we specialise to massless QCD in this section.
Then, integrating eq.(2.4) and evaluating with the sources set to their
physical values (i.e.~zero, apart from the source $\o(x)$ which
becomes the QCD theta angle $\o$), we find
$$
{\pl W\over\pl\o} ~=~ \int d^4 x ~{\d W\over\d\o(x)} ~=~0
\eqno(3.1)
$$
That is, massless QCD is independent of the theta angle.
In fact[10, ch.~23], the same conclusion holds if any of the quark masses 
were to vanish. Then, the theta angle would have no effect and the strong CP
problem would be automatically resolved in QCD. This is, however, an 
unrealistic solution since even $m_u \neq 0$.

It is equally simple to establish that the (zero-momentum) topological 
susceptibility vanishes for massless QCD. This can be read off
immediately from the $M=0$ limit of eq.(2.15). There is, however, one 
subtlety here which is worth noticing. Writing the second identity
in eq.(2.11) in canonical form, we have
$$
ik^\m \langle0|T~J_{\m5}^0~Q|0\rangle - 2n_f\langle0|T~Q~Q|0\rangle = 0
\eqno(3.2)
$$
If there is no massless pseudoscalar meson (e.g.~a $U_A(1)$ Goldstone boson)
coupling to the current, then clearly the first term vanishes at
zero momentum. However, in this case the same conclusion follows even
if there does exist such a particle, since the anomaly equation implies
that the coupling of $Q$ to this massless boson would vanish on-shell.
In either case, we deduce
$$
\chi(0) = \langle0|T~Q~Q|0\rangle\big|_{k=0} = 0
\eqno(3.3)
$$

\vskip0.3cm
The next question is what happens to Goldstone's theorem in the presence
of the anomaly. For the non-anomalous flavour non-singlet currents, the 
third identity in eq.(2.11) shows as usual that (for $M=0$) if there is
a symmetry breaking VEV $\Phi \neq 0$ then there must exist, by Goldstone's
theorem, a massless boson coupling derivatively to the current.
What about the $U_A(1)$ current? The relevant Ward identity, in canonical
form, reads
$$
ik^\m \langle0|T~J_{\m5}^0~\phi_5^0|0\rangle - 
2n_f\langle0|T~Q~\phi_5^0|0\rangle = 2\langle\phi^0\rangle
\eqno(3.4)
$$
Since the r.h.s.~is non-zero, in the absence of the anomaly term
the only way the identity can be satisfied at $k=0$ is if there
exists a massless Goldstone boson coupling to the current.
However, the presence of the extra correlation function involving $Q$
means that this conclusion no longer holds. The Goldstone theorem is
evaded by virtue of the anomaly and there is no physical massless
$U_A(1)$ Goldstone boson.

At its most basic, this is the resolution of the famous $U_A(1)$ problem.
However, things are of course not so simple. Recall from eq.(1.3) that $Q$
may be written as the divergence of the (gauge non-invariant) Chern-Simons
current $K_\m$. We can therefore construct a conserved, but gauge 
non-invariant, current $\hat J_{\m5}^0$ as follows
$$
\hat J_{\m5}^0 = J_{\m5}^0 - 2n_f K_\m
\eqno(3.5)
$$
which satisfies the chiral Ward identity
$$
ik^\m \langle0|T~\hat J_{\m5}^0~\phi_5^0|0\rangle  
= 2\langle\phi^0\rangle
\eqno(3.6)
$$
Applying Goldstone's theorem naively, we would then deduce that after all
there must exist a massless boson in the QCD spectrum.
However, this conclusion is false, although the precise reasons are
still the subject of some debate. A full technical analysis of this
$U_A(1)$ problem would need (at least) another lecture, so instead we
simply refer to the literature[23,24]. In brief, however, there are two
possible escape routes (both of which may be true in a sufficiently
precise formulation of the problem):

\noindent (i)~the boundary conditions at spatial infinity on the 
behaviour of the gauge-variant current $\hat J_{\m5}^0$ imposed by
the vacuum structure of QCD allow eq.(3.6) to be satisfied
without coupling to a massless boson;

\noindent (ii)~a massless boson does indeed exist; however, it decouples
from the positive-norm Hilbert space through a mechanism known as `quartet
decoupling'. This is essentially the same mechanism as is
responsible for the decoupling from the physical spectrum of the ghosts and 
longitudinal and scalar components of the photon field in QED formulated
in a covariant gauge. Its possible application to the $U_A(1)$ problem
has been elaborated by Kugo[25], who shows how the required Goldstone
quartet is constructed by acting with the BRS operator $Q_B$ on the
gauge-variant Goldstone field coupling to $\hat J_{\m5}^0$.
It represents a more realistic generalisation of the popular 
Kogut-Susskind dipole hypothesis[26] and reduces
to it in the one case where the Kogut-Susskind hypothesis has been
proved to work - the 2 dim Schwinger model. In this model, the `quartet'
splits into two `dipoles' which independently cancel from the 
physical spectrum.

\vskip0.3cm
Having convinced ourselves there is no real paradox associated with
the Goldstone theorem applied to the anomalous $U_A(1)$ current, we 
can ask whether it is possible to do better and identify the
mass of the `would-be Goldstone boson' $\eta'$ in some way with
the anomaly. The answer is provided, in the context of the $1/N_c$
approximation to QCD, by the Witten-Veneziano mass formula[11,12].

We know that in nature the lightest pseudoscalar meson in the flavour
singlet channel coupling to the current $J_{\m5}^0$ is the $\eta'$. 
It follows from the second Ward identity in eq.(2.11) that this must
also couple to the operator $Q$, producing a pole at $k^2 = m_{\eta'}^2$
in the topological susceptibility $W_{\o\o}$. However, we have also seen
that $W_{\o\o}$ vanishes at $k=0$. A reasonable parametrisation of the
correlation function is therefore
$$
W_{\o\o}(k^2) = {1\over(2n_f)^2} {k^2\over k^2- m_{\eta'}^2} A(k^2)
\eqno(3.7)
$$
where $A(k^2)$ is a pole-free, and therefore relatively smooth, function
below the glueball threshold. $A(k^2)$ is otherwise unconstrained by
the chiral Ward identities. 

The residue of the pole evidently satisfies
$$
\bigl|\langle0|Q|\eta'\rangle\big|^2 = {1\over(2n_f)^2} m_{\eta'}^2 
A(m_{\eta'}^2)
\eqno(3.8)
$$
and so defining the RG non-invariant `decay constant' $f_{\eta'}$ by
(see refs.[19,21] for a careful discussion of this point) 
$$
\langle0|J_{\m5}^0|\eta'\rangle = ik_\m f_{\eta'}
\eqno(3.9)
$$
and using the anomaly equation, we find
$$
f_{\eta'}^2 m_{\eta'}^2 = A(m_{\eta'}^2)
\eqno(3.10)
$$
Notice that neither side is a RG invariant, each scaling with the
anomalous dimension $2\c$.

We now introduce the large $N_c$ approximation. To leading order in $1/N_c$,
the anomaly is absent and the $\eta'$ mass vanishes. Formally, this
is implemented by assuming that $m_{\eta'}$ is $O(1/N_c)$.
In fact, this can be alternatively re-formulated by using the OZI limit, 
i.e.~the approximation to QCD in which the OZI rule becomes exact.
The OZI limit is precisely defined[27] as the truncation of full QCD in which 
non-planar and quark-loop diagrams are retained, but diagrams in which the 
external currents are attached to distinct quark loops, so that there are
purely gluonic intermediate states, are omitted. (This last fact makes
the connection with the familiar phenomenological form of the
OZI, or Zweig, rule.) This is a more
accurate approximation to full QCD than either the leading large 
$1/N_c$ limit, the quenched approximation (small $n_f$ at fixed $N_c$) 
or the leading topological expansion ($N_c\ra\infty$ at fixed $n_f/N_c$.
In the OZI limit, the $U_A(1)$ anomaly is absent, as is meson-glueball
mixing, and there is an extra $U_A(1)$ Goldstone boson. 

Applying the large $N_c$ limit to the l.h.s.~of eq.(3.7), we therefore
find
$$
\lim_{k\rightarrow 0} \lim_{N_c\rightarrow\infty} W_{\o\o}(k^2) = 
{1\over(2n_f)^2} A(0)
\eqno(3.11)
$$
(Notice that the zero-momentum and large $N_c$ limits do not commute.)
We now put eqs.(3.10) and (3.11) together. To leading order in
$1/N_c$, we may write $f_{\eta'} \simeq \sqrt{2n_f} f_\pi$
and $A(m_{\eta'}) \simeq A(0)$. On the other hand, we may identify the
l.h.s.~of eq.(3.11) as the topological susceptibility in pure
gluodynamics, since as noted above, mixing between the meson and glueball
sectors vanishes to leading order in $1/N_c$, or equivalently in the
OZI limit.

Putting this together, we finally deduce the Witten-Veneziano formula
for the mass of the $\eta'$, valid to leading order in the $1/N_c$
expansion:
$$
m_{\eta'}^2 ~=~ {2n_f\over f_\pi^2} ~\chi^{YM}(0)
\eqno(3.12)
$$
As discussed in the introduction, this is our first example of an explicit
relation linking quark dynamics in the $U_A(1)$ channel to gluon topology
in the form of the topological susceptibility of pure gluodynamics.

\vskip1cm

\noindent{\bf 4.~~Topological Charge Screening and the Ellis-Jaffe
Sum Rule}
\vskip0.5cm

The `proton spin' problem, i.e.~the question of why the first moment of the
flavour singlet component of the polarised proton structure function $g_1^p$ 
is anomalously suppressed, has inspired an impressive research effort, both 
theoretical and experimental, for over a decade. A recent review of this
whole topic from the viewpoint adopted here can be found in ref.[16]. 
As is well-known from standard DIS theory, the first moment of $g_1^p$ can be 
expressed in terms of the axial charges of the proton as follows:
$$
\C_1^p ~\equiv~ \int_0^1 dx~ g_1^p(x;Q^2) ~=~
{1\over12} C_1^{\rm NS}(\a_s)~ \Bigl( a^3 + {1\over3} a^8 \Bigr) +
{1\over9} C_1^{\rm S}(\a_s)~ a^0(Q^2)
\eqno(4.1)
$$
Here, $C_1^{\rm NS}$, $C_1^S$ are the appropriate Wilson coefficients arising 
from the OPE for two electromagnetic currents, while the axial charges are 
defined from the forward matrix elements of the axial currents with the 
normalisations
$$
\langle p,s|J_{\m 5}^3|p,s\rangle = {1\over2} a^3 s_\m  ~~~~~~~
\langle p,s|J_{\m 5}^8|p,s\rangle = {1\over{2\sqrt3}} a^8 s_\m  ~~~~~~~
\langle p,s|J_{\m 5}^0|p,s\rangle = a^0(Q^2) s_\m
\eqno(4.2)
$$
where $s_\m = \bar u(p,s)\c_\m \c_5 u(p,s)$ is the proton polarisation vector.
$a^3$ and $a^8$ are known in terms of the $F$ and $D$ coefficients from
beta and hyperon decay, so that an experimental determination of the
first moment of $g_1^p$ in polarised DIS allows a 
determination of the singlet axial charge $a^0(Q^2)$.
The `proton spin' problem is the fact that it is found experimentally that
$a^0(Q^2)$ is strongly suppressed relative to $a^8$, which would be its 
expected value if the OZI rule were exact in this channel.

\vskip0.2cm
DIS is normally described theoretically using the QCD parton model.
In this model, the axial charges are represented (in the AB renormalisation
scheme[28,29]) in terms of moments of parton distributions as follows[28]:
$$
a^3 = \D u - \D d ~~~~~~~~
a^8 = \D u + \D d - 2\D s ~~~~~~~~
a^0(Q^2) = \D u + \D d + \D s - n_f {\a_s\over2\pi} \D g(Q^2)
\eqno(4.3)
$$
In the parton model, the `proton spin' problem takes the following 
form. In the naive, or valence quark, parton model we would expect the
strange quark and gluon distributions to vanish, i.e.~$\D s = 0$,
$\D g(Q^2) = 0$. In that case, $a^0 = a^8$, the OZI prediction.
Inserted into eq.(4.1), this gives the Ellis-Jaffe sum rule[2]. However,
the observed suppression $a^0(Q^2) < a^8$ can be accommodated in the 
full QCD parton model by invoking either or both a non-zero polarised 
strange quark distribution $\D s \neq 0$ or a non-zero polarised gluon
distribution $\D g(Q^2) \neq 0$. An interesting conjecture (in line with
the insights of the approach discussed here) is that the suppression 
is primarily due to the gluon distribution[28], although a quantitative 
prediction would still only follow if $\D g(Q^2)$ can be independently 
measured, either through a precise analysis of the $Q^2$ dependence of 
$g_1^p$ [29] or directly through other less inclusive high energy processes 
such as open charm production.
Notice, however, that even in the QCD parton model picture, it is {\it not}
possible to identify $a^0(Q^2)$ with spin[17,16]. This identification only 
holds for free quarks, in which case the $Q^2$ scale dependence, which is 
related through eq.(2.9) (see also eq.(5.26)) to the $U_A(1)$ anomaly, 
disappears from $a^0$. It must be emphasised that the so-called 
`proton spin' problem is not a problem of {\it spin} -- rather, it is a 
question of understanding the dynamical origin of the OZI violation 
$a^0(Q^2) < a^8$. 

\vskip0.2cm
In this section, we shall discuss a less conventional approach (the `CPV'
method[16]) to DIS based on the composite operator propagator -- proper vertex 
formalism described in section 2. The starting point, as indicated above, 
is the use of the OPE in the proton matrix element of two currents.
This gives the standard form for a generic structure function moment:
$$
\int_0^1 dx~ x^{n-1} F(x;Q^2) = \sum_i C_i^n(Q^2) \langle p|\OO_i^n(0)
|p\rangle
\eqno(4.4)
$$
where $\OO_i^n$ are the set of lowest twist, spin $n$ operators in the OPE
and $C_i^n(Q^2)$ the corresponding Wilson coefficients. 
In the CPV approach, we now factorise the matrix element into the product 
of composite operator propagators and vertex functions, as illustrated in
Fig.~1.
\vskip0.2cm
\centerline{
{\epsfxsize=3.5cm\epsfbox{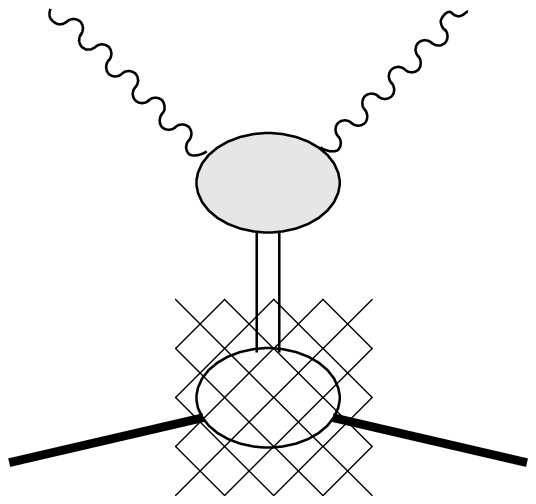}}}
\noindent{\eightrm Fig.1~~CPV description of DIS. The double line denotes
the composite operator propogator and the lower blob the 1PI vertex.}
\vskip0.2cm
To do this, we first select a set of composite operators $\tilde {\OO}_i$
appropriate to the physical situation and define vertices 
$\C_{\tilde{\OO}_i pp}$ as 1PI with respect to this set.
Technically, this is achieved as in eq.(2.16) by introducing sources for 
these operators in the QCD generating functional, then performing a 
(partial) Legendre transform to obtain a generating functional 
$\C[\tilde{\OO}_i]$. The 1PI vertices are the functional derivatives 
of $\C[\tilde{\OO}_i]$. The generic structure function sum rule (4.4) 
then takes the form
$$\eqalignno{
\int_0^1dx~x^{n-1}~F(x,Q^2) &=  
\sum_i \sum_j C_j^{(n)}(Q^2) \langle0|T~\OO_j^{(n)} ~
\tilde{\OO}_i |0\rangle \C_{\tilde{\OO}_i pp} \cr
&= \sum_i \sum_j C_j P_{ji} V_i
&(4.5) \cr }
$$
in a symbolic notation.

This decomposition splits the structure function into three pieces -- first,
the Wilson coefficients $C_j^{(n)}(Q^2)$ which control the $Q^2$ dependence 
and can be calculated in perturbative QCD; second, non-perturbative but
{\it target-independent} QCD correlation functions 
$\langle0|T~\OO_j^{(n)} ~\tilde{\OO}_i |0\rangle$; and third,
non-perturbative, target-dependent vertex functions $\C_{\tilde{\OO}_i pp}$
describing the coupling of the target proton to the composite operators 
of interest.
The vertex functions cannot be calculated directly from first principles.
They encode the information on the nature of the proton state and play an 
analogous role to the parton distributions in the more conventional
parton picture. 

It is important to recognise that this
decomposition of the matrix elements into products of propagators
and proper vertices is {\it exact}, independent of the choice of
the set of operators $\tilde{\OO}_i$. In particular, it is not necessary
for $\tilde{\OO}_i$ to be in any sense a complete set. All that happens if a 
different choice is made is that the vertices $\C_{\tilde{\OO}_i pp}$
themselves change, becoming 1PI with respect to a different
set of composite fields. Of course, while any set of $\tilde{\OO}_i$ may be
chosen, some will be more convenient than others. Clearly, the set 
of operators should be as small as possible while still capturing the
essential physics (i.e.~they should encompass the relevant degrees of
freedom) and indeed a good choice can result in vertices $\C_{\tilde{\OO}_i pp}$
which are both RG invariant and closely related to low energy physical 
couplings, such as $g_{\p NN}$. In this case, 
eq.(4.5) provides a rigorous relation between high $Q^2$ DIS and low-energy 
meson-nucleon scattering. 

For the first moment sum rule for $g_1^p$~[19,7,16], it is most convenient 
to use the $U_A(1)$ anomaly equation immediately to re-express $a^0(Q^2)$ 
in terms of the forward matrix element of the topological charge $Q$, i.e.
$$
a^0(Q^2) ~=~{1\over 2m_N} 2n_f \langle p|Q|p\rangle
\eqno(4.6)
$$
where $m_N$ is the nucleon mass.

Our set of operators $\tilde{\OO}_i$ is then chosen to be the renormalised 
flavour singlet pseudoscalars $Q$ and $\Phi_5$, where 
$\Phi_5$ is simply the operator $\phi_5^0$ of eqs.(2.1) and (2.9)
with a special, and crucial, normalisation.
The normalisation factor is chosen such that in the absence of the 
anomaly, or more precisely in the OZI limit of QCD (see section 3), 
$\Phi_{5}$ would have the correct normalisation to couple with unit decay 
constant to the $U_A(1)$ Goldstone boson which would exist in this limit. 
This also ensures that the vertex is RG scale independent, as we prove in
section 5 where we relate this discussion to the $U_A(1)$ 
Goldberger-Treiman relation. The vertices are defined from the 
generating functional (2.16).
We then have
$$
\C_{1~singlet}^p ={1\over9} {1\over2m_N} 2n_f
C_1^{\rm S}(\a_s)   
\biggl[\langle 0|T~Q~ Q|0\rangle \hat\C_{Qpp}
+\langle 0|T~ Q~ \Phi_{5}|0\rangle \hat\C_{\Phi_5 pp} \biggr]
\eqno(4.7)
$$
where the propagators are at zero momentum and the vertices 
are 1PI wrt $Q$ and $\Phi_{5}$ only. 
For simplicity, we have also introduced the notation
$i\bar u \C_{Qpp}u = \hat\C_{Qpp} \bar u \c_5 u$, etc.
This is illustrated in Fig.~2.
\vskip0.3cm
\centerline{
{\epsfxsize=8cm\epsfbox{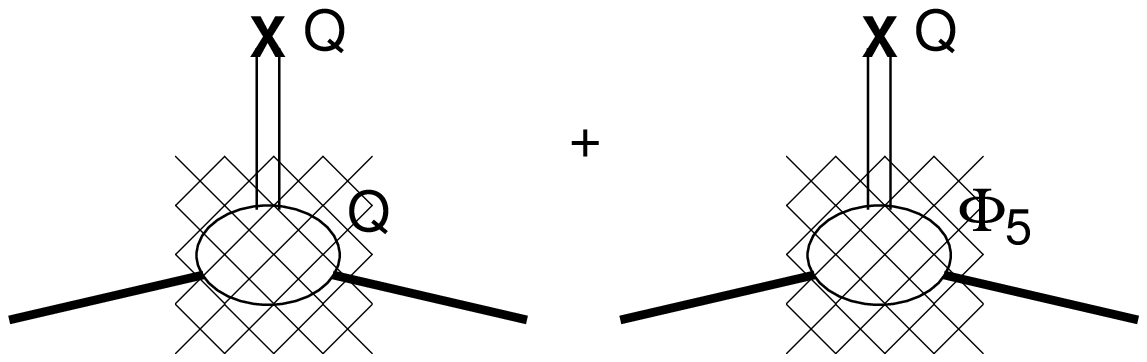}}}
\vskip0.3cm
\noindent{\eightrm Fig.2~~CPV decomposition of the matrix element 
$\langle p|Q|p\rangle$.}
\vskip0.2cm
The composite operator propagator in the first term is simply 
the (zero-momentum) QCD topological susceptibility $\chi(0)$ which,
as we have seen in sections 2 and 3, vanishes for QCD with massless quarks.
Furthermore, with the normalisation specified above for $\Phi_5$, the  
propagator $\langle 0|T~ Q~\Phi_5|0\rangle$ at zero momentum
is simply the square root of the slope of the topological
susceptibility. 

To see this, notice that by virtue of their definition in terms of
the generating functional (2.16), the matrix of 2-point vertices in the 
pseudoscalar sector is simply the inverse of the corresponding matrix
of pseudoscalar propagators, i.e.
$$
\left(\matrix{\C_{QQ} &\C_{Q\phi_5^0} \cr 
\C_{\phi_5^0 Q} &\C_{\phi_5^0 \phi_5^0} \cr }\right) ~=~ -
\left(\matrix{W_{\o\o} & W_{\o S_5^0} \cr
W_{S_5^0\o} & W_{S_5^0 S_5^0} \cr }\right)^{-1}
\eqno(4.8)
$$
This implies 
$$
\C_{\phi_5^0 \phi_5^0} = - W_{\o\o} \bigl({\rm det}~
W_{{\cal S}{\cal S}}\bigr)^{-1}
\eqno(4.9)
$$
letting ${\cal S}$ represent the set $\{\o, S_5^0\}$.
Differentiating w.r.t.~$k^2$ and taking the limit $k^2 = 0$, and 
exploiting the fact that $W_{\o\o}(0)$ vanishes, we find
$$
{d\over dk^2}\C_{\phi_5^0 \phi_5^0}\big|_{k=0} =
\chi'(0) W_{\o S_5^0}^{-2}(0)
\eqno(4.10)
$$
Finally, normalising the field $\Phi_5$ proportional to $\phi_5^0$ such 
that ${d\over dk^2}\C_{\Phi_5 \Phi_5}\big|_{k=0} = 1$ (see also
the discussion in section 5, following eq.(5.7)), we find the 
required relation
$$
\langle0|T~Q~\Phi_5|0\rangle\big|_{k=0} = \sqrt{\chi'(0)}
\eqno(4.11)
$$

We therefore find:
$$
\C_{1~singlet}^p ~=~ {1\over9} {1\over 2m_N} 2n_f~
C_1^{\rm S}(\a_s) ~\sqrt{\chi^{\prime}(0)} ~\hat\C_{\Phi_{5} pp}
\eqno(4.12)
$$
The slope of the topological susceptibility $\chi'(0)$ is
not RG invariant but, as shown in eq.(2.21), scales with the anomalous 
dimension $2\c$, i.e.
$$
{d\over dt} \sqrt{\chi'(0)} = \c \sqrt{\chi'(0)}
\eqno(4.13)
$$
On the other hand, the proper vertex has been chosen specifically
so as to be RG invariant, as proved in section 5. The renormalisation group 
properties of this decomposition are crucial to our proposed resolution 
of the `proton spin' problem.
 
Our proposal[19] is that we should expect the source of OZI 
violations to lie in the RG non-invariant, and therefore anomaly-sensitive, 
terms, i.e. in $\chi^{\prime}(0)$ rather than in the RG invariant vertex.
Notice that we are using RG non-invariance, i.e.~dependence 
on the anomalous dimension $\c$, merely as an indicator of which quantities
are sensitive to the anomaly and therefore likely to show OZI violations.
Since the anomalous suppression in $\C_1^p$ is thus assigned to the composite 
operator propagator rather than the proper vertex, the suppression is a 
{\it target independent} property of QCD related to the anomaly, 
not a special property of the proton structure.
  
To convert this into a quantitative prediction we use the OZI 
approximation for the vertex  $\hat\C_{\Phi_{5} pp}$. In terms of a 
similarly normalised octet field $\Phi_5^8$, this is
$\hat\C_{\Phi_5 pp} = \sqrt2 \hat\C_{\Phi_5^8 pp}$. 
Notice that the normalisation is crucial in allowing the use of the
OZI relation here. The corresponding OZI 
prediction for $\sqrt{\chi'(0)}$ would be $f_{\pi}/\sqrt6$. These OZI 
values are determined by comparing the result (4.12) (at least that part
relating to the proton matrix element) with the conventional
Goldberger-Treiman relation for the flavour octet axial charge in 
the chiral limit (see section 5). This gives our formula 
$$
{a^0(Q^2)\over a^8} = {\sqrt6\over f_{\pi}} \sqrt{\chi'(0)}\big|_{Q^2}
\eqno(4.14)
$$
for the flavour singlet axial charge. Incorporating this into
the formula for the first moment of the polarised structure function, 
we find
$$
\C_{1~singlet}^p ~=~ {1\over9} C_1^{\rm S}(\a_s) ~a^8~
{\sqrt6\over f_{\pi}}~\sqrt{\chi'(0)}
\eqno(4.15)
$$

Finally, substituting this into the expression for the complete
first moment and using our spectral sum rule derivation of $\chi'(0)$,
which gives a suppression factor of around $0.6$ (see refs.[7,8,16] 
for more details), we obtain our prediction
in the chiral limit
$$\eqalignno{
a^0(Q^2=10\GV^2) &= 0.33 \pm 0.05 \cr
\C_1^p(Q^2=10\GV^2) &= 0.144 \pm 0.009
&(4.16) \cr}
$$

This is to be compared with the OZI (Ellis-Jaffe) prediction 
$a^0 = 0.58\pm0.03$ and the current experimental data from the SMC
collaboration[30]:
$$
\C_1^p (Q^2=10\GV^2)\big|_{(x>0.003)} \equiv \int_{0.003}^1 dx~g_1^p(x;Q^2)
= 0.141 \pm 0.012
\eqno(4.17)
$$
The result for the entire first moment depends on how the extrapolation to
the unmeasured small $x$ region $x<0.003$ is performed. This is still
a controversial issue. Using a simple
Regge fit, SMC find $\C_1^p = 0.142 \pm 0.017$ from which they deduce
$a^0= 0.34 \pm 0.17$, while using a small $x$ fit using perturbative QCD 
evolution of the parton distributions[31] they find 
$\C_1^p = 0.130 \pm 0.017$ and
$a^0 = 0.22 \pm 0.17$ ~(all at $Q^2 = 10\GV^2$). 

More recently, SMC have published an alternative analysis[32] of their data,
this time quoting a slightly lower number for the integral over
the measured region of $x$:
$$
\C_1^p (Q^2=10\GV^2)\big|_{(x>0.003)} \equiv \int_{0.003}^1 dx~g_1^p(x;Q^2)
= 0.133 \pm 0.009
\eqno(4.18)
$$

Clearly it is premature to draw too strong a conclusion given the
large errors on the experimental determinations of $\C_1^p$ and
$a^0$ and the uncertainty over the small $x$ extrapolation.
Nevertheless, it is extremely encouraging that our prediction
is firmly in the region favoured by the data and gives us confidence
that our explanation of the `proton spin' problem in terms of the
topological susceptibility is correct. (For a more extensive discussion
of the phenomenological aspects of this approach to the `proton spin'
problem, and a proposal to test our `target-independence' conjecture
in semi-inclusive DIS experiments at polarised HERA, see refs.[16,33,34]).

Finally, we should draw attention to the basic dynamical mechanism
responsible for the suppression of the `proton spin'. What we have
shown is that when a matrix element of the topological charge is
measured, the QCD vacuum screens the topological charge through
the zero or anomalously small values of the susceptibility
$\chi(0)$ and its slope $\chi'(0)$ respectively (see Fig.~2).
The mechanism is analogous to the screening of electric charge in QED.
There, because of the gauge Ward identity, the screening is given entirely
by the (`target independent') dressing of the photon propagator by
vacuum polarisation diagrams, leading to the relation $e_R = e_B
\sqrt{Z_3}$ (with $Z_3<1$) between the renormalised and bare charges,
in direct analogy to eq.(4.14) above with the topological susceptibility
playing the role of the photon propagator.

\vskip1cm

\noindent{\bf 5.~~The $U_A(1)$ Goldberger-Treiman Formula\footnote{${}^2$}
{\eightpoint 
\noindent The material in this section was not presented in the lecture
in Zuoz. However, it is closely related to the topics described
in sections 2-4 and should help to clarify the discussion given there.}}
\vskip0.5cm
In this final section, we present a unified derivation of the 
Goldberger-Treiman relations (1.9) for the flavour singlet and 
non-singlet axial charges for QCD with non-vanishing quark masses
in the functional composite operator propagator -- vertex formalism[8]. 
The zero-mass limit of the formula for the singlet $U_A(1)$ charge
reduces to the expression already used in section 4 (see eq.(4.12)) in the 
context of the `proton spin' problem. Notice that these
new GT relations are more general than the conventional formulae.
Indeed, they are {\it exact} field-theoretic results in QCD.
The familiar PCAC forms (in the flavour non-singlet channels)
are obtained by approximating the 1PI vertices by the corresponding
low-energy meson-nucleon couplings and by approximating the slopes
of the current correlation functions (1.10) by decay constants. Away from 
the chiral limit, both these approximations assume pole dominance of the 
matrix elements and correlation functions by the pseudo-Goldstone bosons.

The axial charges $G_A^a$ are defined as the form factors in the forward
nucleon matrix elements of the axial currents, viz. 
$$
\langle p,s|J_{\m5}^a|p,s\rangle = G_A^a s_\m 
\eqno(5.1)
$$
(Compare with the normalisations in eq.(4.2).)

To express this matrix element in terms of composite operator propagators
and the associated 1PI vertices, we first introduce an interpolating field 
$N$ and source $S_N$ for the nucleon. (Notice that this is purely a 
formal device -- there is no dynamics implicit in this step.) 
The matrix element is then just the 3-point function $W_{V_{\m5}^a S_N S_N}$ 
with the external propagators amputated. This can be re-expressed in terms of
the vertex functional $\C$ as follows (see refs.[19,8] for a derivation of the
relevant formulae involving partial Legendre transforms):
$$\eqalignno{
\langle p,s|J_{\m5}^a|p,s\rangle ~&=~ \bar u(p,s) \biggl[
W_{S_N S_N}^{-1}~ W_{V_{\m5}^a S_N S_N}~ W_{S_N S_N}^{-1}\biggr] 
u(p,s) \cr &{}\cr
&= ~\bar u(p,s)\biggl[\C_{V_{\m5}^a NN} ~+~ W_{V_{\m5}^a \o}~ 
\C_{QNN}~ +~ W_{V_{\m5}^a S_5^b}~ \C_{\phi_5^b NN}\biggr] u(p,s) 
&(5.2) \cr }
$$
Since the propagators on the r.h.s.~vanish at zero momentum (this requires 
the absence of any $1/k^2$ poles, which as we have seen in section 3 
is assured by the $U_A(1)$ anomaly and quark masses), we find simply
$$
2m_N G_A^a~\bar u \c_5 u = \bar u \biggl[k_\m \C_{V_{\m5}^a NN} \big|_{k=0}
\biggr] u 
\eqno(5.3)
$$
The GT relations then follow immediately from the Ward identity (2.17) 
for $\C$. Differentiating w.r.t.~the nucleon fields, we find
$$
2m_N G_A^a =  \Phi_{ab} \hat\C_{\phi_5^b NN} \big|_{k=0} 
\eqno(5.4)
$$
where as before we define $i\bar u ~\C_{\phi_5^a NN}~ u = 
\hat\C_{\phi_5^a NN} ~\bar u \c_5 u$, etc.

A non-forward version of the GT relation, which is closer to the
analysis in section 4, can also be found from eq.(5.2) by using the 
Ward identities (2.11) for the propagators together with (2.17) 
for $\C_{V_{\m5}^a NN}$.
This allows us to write, for all $k$,
$$\eqalignno{
2m_N G_A^a(k^2) + k^2 G_P^a(k^2) = &- \Bigl(2n_f \d_{a0} W_{\o\o} + M_{ac}
W_{S_5^c\o}\Bigr) \hat\C_{QNN} \cr
&- \Bigl(2n_f \d_{a0} W_{\o S_5^b} + M_{ac} W_{S_5^c S_5^b}\Bigr)
\hat\C_{\phi_5^b NN}
&(5.5) \cr }
$$
$G_P^a(k^2)$ is the pseudoscalar form factor in the non-forward matrix
element $\langle p,s|J_{\m5}^a|p,s\rangle$, and again has no $1/k^2$ pole
for the reasons given above. This expression clearly reduces to eq.(5.4) on
using the zero-momentum Ward identities (2.14) for the propagators.
In the chiral limit ($M=0$), this can be compared with eq.(4.7).

The remaining step to convert eq.(5.4) into the useful form of the GT
relations is to normalise the field $\phi_5^a$ appropriately. Clearly,
eq.(5.4) is independent of the normalisation. However, with a suitable
choice, the vertices can be made both RG invariant and essentially
identical to the physical Goldstone boson couplings $g_{\pi NN}$ etc. To
achieve this, we define normalised fields 
$$
\eta^a = B_{ab} \phi_5^b
\eqno(5.6)
$$
where $B$ is a constant matrix such that\footnote{${}^3$}{\eightpoint
\noindent Applied to the flavour singlet field $\phi_5^0$ in the chiral
limit, this is the same condition used to normalise the field $\Phi_5$
in section 4.} 
$$
{d\over dk^2} \C_{\eta^a \eta^b} \big|_{k=0} = \d_{ab} 
\eqno(5.7)
$$
This condition ensures that the fields $\eta^a$ have unit coupling to the
Goldstone bosons. 

The case of the singlet $\eta^0$ is of course special, since it is only
after mixing with the topological field $Q$ (and then flavour mixing)
that it becomes the physical $\eta'$. In fact, this is why it is most 
convenient to impose the normalisation condition as
above on the matrix of 2-point vertices $\C_{\eta^a \eta^b}$, which is the
inverse of the pseudoscalar propagator matrix, since this most simply
characterises the $\eta^0$ before mixing with $Q$. The corresponding state
is the unphysical `OZI Goldstone boson' introduced in ref.[17] and
extensively discussed in refs.[19,21].

It can now be proved that the vertices $\hat\C_{\eta^a NN}$ defined with the 
fields normalised according to eq.(5.7) are RG invariant. The proof
is based on the functional form of the RGEs introduced in section 2
and is given at the end of this section. 

Re-expressing eq.(5.4) in terms of the properly normalised vertices, we have
$$
2m_N G_A^a =  \Phi_{ac} B_{cb}^T ~\hat\C_{\eta^b NN}\big|_{k=0} 
\eqno(5.8)
$$
where $B$ is to be determined from the 2-point vertex condition 
$$
{d\over dk^2} \C_{\phi_5^a \phi_5^b}\big|_{k=0} = B_{ac}^T ~{d\over
dk^2}\C_{\eta^c \eta^d}\big|_{k=0} ~B_{db} = B_{ac}^T B_{cb}
\eqno(5.9)
$$
The straightforward approach to finding $\C_{\phi_5^a \phi_5^b}$ is to
write it as one component of the inverse of the propagator matrix 
$W_{{\cal S} {\cal S}}$ (with ${\cal S} = \{\o,S_5^a\}$). 
This was the approach used in section 4 to relate 
$\C_{\phi_5^0 \phi_5^0}$ to the topological
susceptibility $W_{\o\o}$. However, inverting this matrix in the
$m_q\neq0$, multi-flavour case is cumbersome, so here we use an 
indirect but more elegant method.
First, we use the Ward identity (2.19) to express $\C_{\phi_5^a
\phi_5^b}$ in terms of $\C_{V_{\m5}^a V_{\n5}^b}$:
$$
k_\m k_\n \C_{V_{\m5}^a V_{\n5}^b} ~=~
\Phi_{ac} ~\C_{\phi_5^c \phi_5^d} ~\Phi_{db}~ - ~M_{ac} \Phi_{cb}
\eqno(5.10)
$$
Then, using a general identity for partial Legendre transforms[19,8], we
relate $\C_{V_{\m5}^a V_{\n5}^b}$ to the current propagator $W_{V_{\m5}^a
V_{\n5}^b}$:
$$
\C_{V_{\m5}^a V_{\n5}^b} ~=~ W_{V_{\m5}^a V_{\n5}^b}~ - ~W_{V_{\m5}^a {\cal
S}} ~W_{{\cal S} {\cal T}}^{-1}~ W_{{\cal T} V_{\n5}^b}
\eqno(5.11)
$$
where ${\cal S}$ and ${\cal T}$ represent $\{\o,S_5^a\}$. Combining with
eqs.(2.12) and (2.19) finally yields $$
\Phi_{ac}~ \C_{\phi_5^c \phi_5^d}~ \Phi_{db} ~= ~ W_{S_D^a S_D^b} ~+~ k_\m
W_{V_{\m5}^a {\cal S}} ~W_{{\cal S} {\cal T}}^{-1}~ W_{{\cal T} V_{\n5}^b}
k_\n
\eqno(5.12)
$$
and so, in matrix notation,
$$
\Phi B^T B \Phi ~=~ {d\over dk^2} W_{S_D S_D}\big|_{k=0}~ +~ {d\over
dk^2}\bigl(k_\m W_{V_{\m5} {\cal S}} ~W_{{\cal S} {\cal T}}^{-1}~ W_{{\cal
T} V_{\n5}} k_\n\bigr)\big|_{k=0} 
\eqno(5.13)
$$
The argument is almost complete. $\Phi B^T$ is precisely the combination we
need for the GT relations (5.8), and is related by eq.(5.13) directly to
the first moment of the correlation function (2.13) for the
divergences of the currents. 

It remains to show that the final term in eq.(5.13) vanishes. The first and
last factors are of $O(k^2)$, so this will contribute zero unless there is
a $1/k^2$ pole in the inverse pseudoscalar propagator matrix $W_{{\cal S}
{\cal T}}^{-1}$. As already mentioned, there are no $1/k^2$ poles in the
propagators themselves, so all we need show is that the determinant
$\D(k^2)$ of this propagator matrix is non-vanishing at $k=0$. This follows
from the formula
$$
\D(0) = W_{\o\o}(0)~ ({\rm det} M)^{-1}~ {\rm det} \Phi 
\eqno(5.14)
$$
since ${\rm det} M$ and ${\rm det} \Phi$ are non-zero (eqs.(2.8)) and
$W_{\o\o}$ is non-vanishing away from the chiral limit. An elegant proof of
eq.(5.14), based on the zero-momentum Ward identities for $W_{{\cal S}
{\cal T}}$ is as follows. Define
$$
\hat M = \left(\matrix{{\bf 1} &0 \cr 0 &M \cr}\right) ~~~~~~~~~~~~~
W = \left(\matrix{
W_{\o\o}&W_{\o S_5^b} \cr W_{S_5^a \o} &W_{S_5^a S_5^b}}\right) 
\eqno(5.15)
$$
with $\D = {\rm det} W$. Then,
$$\eqalignno{
{\rm det} M ~~\D &= {\rm det} \hat M ~{\rm det} W \cr &{}\cr
&= \left|\matrix{
W_{\o\o}&W_{\o S_5^b} \cr M_{ac} W_{S_5^c \o} &M_{ac} W_{S_5^c
S_5^b}}\right| 
&(5.16) \cr }
$$
Using the zero-momentum Ward identities (2.14) and, in the case of the
$a=0$ row taking a linear combination with the first ($\o$) row, the
determinant simplifies, leaving
$$\eqalignno{
{\rm det} M ~~\D &= \left|\matrix{
W_{\o\o}&W_{\o S_5^b} \cr 0 &-\Phi_{ab}}\right| \cr &{}\cr
&= W_{\o\o} ~{\rm det} \Phi
&(5.17) \cr }
$$
as required.

This completes the proof of the GT relations. To summarise, we have shown
that the flavour singlet and non-singlet axial charges are given by the
single, unified relation 
$$
2m_N G_A^a ~=~  F_{ab} ~\hat \C_{\eta^b NN}\big|_{k=0} 
\eqno(5.18)
$$
where $F \equiv \Phi B^T$ is determined from 
$$
FF^T = {d\over dk^2} W_{S_D S_D}\big|_{k=0} 
\eqno(5.19)
$$
that is,
$$
F_{ac} F_{cb}^T = \lim_{k=0} {d\over dk^2}~ i \int dx~e^{ikx}~\langle 0|
T~\pl^\m J_{\m5}^a(x)~\pl^\n J_{\n5}^b(0)|0\rangle
\eqno(5.20)
$$

The current algebra approximation to eq.(5.20) is easily found.
Consider just the $G_A^3$ relation for simplicity, since in this case
isospin invariance (assume $m_u = m_d = 0$) ensures that there is no 
flavour mixing. Applying pole dominance to the current correlator,
we find 
$$
F_{33}^2 = \lim_{k=0} {d\over dk^2} \Bigl(f_\pi^2 k^4 {1\over k^2}\Bigr)
= f_{\pi}^2
\eqno(5.21)
$$
where in the intermediate step we have distinguished the factors of $k^2$
arising from the decay constant and pole terms. 
Under the same PCAC assumptions, the vertex can be identified with the 
low-energy pion-nucleon coupling constant,
i.e. $\hat\C_{\eta^3 NN}|_{k=0} = g_{\pi NN}$.
Writing $G_A^3$ in standard notation as $g_A/2$, where $g_A$ is the 
isotriplet axial-vector coupling of the nucleon measured in beta decay,
we therefore recover in this approximation the standard GT formula
$$
f_\pi g_{\pi NN} = m_N g_A
\eqno(5.22)
$$
We emphasise again, however, that the relations (5.18) and (5.20) are 
{\it exact} -- they are neither dependent on this interpretation nor 
make any PCAC assumption or approximation.

The GT relation simplifies in the chiral limit, where flavour mixing 
is absent. In this case, the singlet axial charge is simply given by 
$$
2m_N G_A^0 = 2n_f~ \sqrt{\chi'(0)}~ \hat \C_{\eta^0 NN}\big|_{k=0} 
\eqno(5.23)
$$
where as usual $\chi'(0)$ is the slope of the topological 
susceptibility. 

This formula can now be recognised as the basis of our resolution of the
`proton spin' problem (see eq.(4.12)).
The general formula (5.18) extends this to include mixing with the flavour 
non-singlet sector, introducing a matrix structure which replaces the 
simple square root of $\chi'(0)$ in eq.(5.23) and generalising the fields 
in the correlation function to the entire divergence of the current including 
mass terms as well as the anomaly $Q$.

\vskip0.3cm
The final task is to determine the RG properties of the various quantities
entering into the derivation of the GT relations, and verify the assertion
that the 1PI vertices defined here are RG invariant.

The RGE for the matrix $B_{ab}$, which relates the
$\phi_5^a$ fields to the canonically normalised $\eta^a$ fields by
$\eta^a = B_{ab}\phi_5^b$, is readily found using the RGE (2.23)
for the 2-point vertex $\C_{\phi_5^a \phi_5^b}$.
From the definition (5.9), using eq.(2.23) and the zero-momentum limit 
of the Ward identities (2.18), we deduce
$$
\DD B_{ab} = -\c_\phi B_{ab} + \c B_{ac} \Phi_{c0} \Phi_{0b}^{-1}
\eqno(5.24)
$$

The RGE for $F_{ab}$ now follows immediately from its definition
$F = \Phi B^T$ and the RGE $\DD \Phi = \c_\phi \Phi$.
It is simply
$$
\DD F_{ab} = \c \d_{a0} F_{0b}
\eqno(5.25)
$$

The final step in proving RG consistency of the unified GT formulae
is to show that the vertices $\hat \C_{\eta^a NN}$ (at ${k=0}$) 
are RG invariant.
Eq.(5.25) then ensures the required RGE for the axial charges, 
$$
\DD G_A^a = \c \d_{a0} G_A^a
\eqno(5.26)
$$
showing that only the singlet axial charge has a non-trivial RG scaling.
To check this explicitly, notice that eq.(2.24) for $\hat \C_{\phi_5^a NN}$
simplifies at $k=0$. The contact terms vanish and using the zero-momentum
Ward identities (see eq.(2.18)) we find
$$
\DD \hat\C_{\phi_5^a NN}\big|_{k=0} = 
-\c_\phi \hat\C_{\phi_5^a NN}\big|_{k=0}
+ \c \Phi_{a0}^{-1} \Phi_{0b} \hat\C_{\phi_5^b NN}\big|_{k=0}
\eqno(5.27)
$$
Since $\hat\C_{\phi_5^a NN}\big|_{k=0} = 
B_{ab}^T \hat \C_{\eta^b NN}\big|_{k=0}$, and comparing eq.(5.27)
with the RGE (5.24) for $B$, we confirm
$$
\DD \hat\C_{\eta^a NN}\big|_{k=0} = 0 ~~~~~~~~{\rm for~ all}~a
\eqno(5.28)
$$

\vskip1cm

\noindent{\bf Acknowledgements}
\vskip0.5cm
I am very grateful to all the organisers, in particular
Dirk Graudenz, for the invitation to lecture at Zuoz and for 
creating such an enjoyable summer school.
I would also like to thank Gabriele Veneziano for a stimulating and
long-standing collaboration on the original research presented here.
This work is partially supported by the EMC TMR Network
Grant FMRX-CT96-0008 and by PPARC.

\vfill\eject

\noindent{\bf References}
\vskip0.5cm

\settabs\+\ [&1] &Author \cr

\+\ [&1] &S.L.~Adler, {\it Phys.~Rev.} 177 (1969) 2426; \cr
\+\ &{}  &J.S.~Bell and R.~Jackiw, {\it Nuovo Cim.} 60 (1969) 47.\cr
\+\ [&2] &J.~Ellis and R.L.~Jaffe, {\it Phys.~Rev.} D9 (1974) 1444.  \cr
\+\ [&3] &A.~Di Giacomo, {\it Nucl.~Phys. (Proc.~Suppl.)} 23A (1991) 191.\cr
\+\ [&4] &G.~Boyd, B.~All\'es, M.~D'Elia and A.~Di Giacomo, {\it in} Proceedings,
HEP 97 Jerusalem, \cr
\+\ &{} &~~~~~hep-lat/9711025.  \cr
\+\ [&5] &A.~Di Giacomo, {\it in} Proceedings, Ahrenshoop Symposium, 1997,
hep-lat/9711034. \cr
\+\ [&6] &S.~Narison, {\it Phys.~Lett.} B255 (1991) 101. \cr
\+\ [&7] &S.~Narison, G.M.~Shore and G.~Veneziano, {\it Nucl.~Phys.}
B433 (1995) 209.\cr 
\+\ [&8] &S.~Narison, G.M.~Shore and G.~Veneziano, hep-ph/9812333. \cr 
\+\ [&9] &E.V.~Shuryak and J.J.M.~Verbaarschot, {\it Phys.~Rev.} D52 (1995)
295. \cr
\settabs\+ [1&1] &Authors \cr
\+ [1&0] &S.~Weinberg, `The Quantum Theory of Fields', Cambridge University 
Press, 1996. \cr
\+ [1&1] &E.~Witten, {\it Nucl.~Phys.} B156 (1979) 269. \cr
\+ [1&2] &G.~Veneziano, {\it Nucl.~Phys.} B159 (1979) 213. \cr
\+ [1&3] &P.~Di Vecchia and G.~Veneziano, {\it Nucl.~Phys.} B171 (1980) 253.\cr
\+ [1&4] &P.~Herrerra-Sikl\'ody, J.I.~Latorre, P.~Pascual and J.~Taron,
 {\it Nucl.~Phys.} B497 (1997) 345. \cr
\+ [1&5] &H.~Leutwyler, {\it Nucl.~Phys. (Proc.~Suppl.)} B64 (1998) 223. \cr 
\+ [1&6] &G.M.~Shore, {\it Nucl.~Phys. (Proc.~Suppl.)} B64 (1998) 167. \cr
\+ [1&7] &G.~Veneziano, {\it Mod.~Phys.~Lett.} A4 (1989) 1605. \cr 
\+ [1&8] &G.M.~Shore and G.~Veneziano, {\it Phys.~Lett.} B244 (1990) 75. \cr 
\+ [1&9] &G.M.~Shore and G.~Veneziano, {\it Nucl.~Phys.} B381 (1992) 23. \cr 
\+ [2&0] &D.~Espriu and R.~Tarrach, {\it Z.~Phys} C16 (1982) 77. \cr
\+ [2&1] &G.M.~Shore and G.~Veneziano, {\it Nucl.~Phys.} B381 (1992) 3. \cr 
\+ [2&2] &G.M.~Shore, {\it Nucl.~Phys.} B362 (1991) 85. \cr
\+ [2&3] &G.~'t Hooft, {\it Phys.~Rev.~Lett.} 37 (1976) 8;
{\it Phys.~Rep.} 142 (1986) 357. \cr
\+ [2&4] &R.J.~Crewther, {\it in} Facts and Prospects of Gauge Theories, 
Schladming 1978,  \cr
\+\ &{}  &~~~~~{\it Acta Phys.~Austriaca Suppl.} XIX (1978)
47; {\it Riv.~Nuovo Cim.} 2 (1979) 63 \cr
\+ [2&5] &T.~Kugo, {\it Nucl.~Phys.} B155 (1979) 368.  \cr
\+ [2&6] &J.~Kogut and L.~Susskind, {\it Phys.~Rev.} D11 (1975) 3594. \cr
\+ [2&7] &G.~Veneziano, {\it in} `From Symmetries to Strings: Forty Years of
Rochester Conferences', \cr
\+\ &{} &~~~~~ed. A.~Das, World Scientific, 1990. \cr 
\+ [2&8] &G.~Altarelli and G.G.~Ross, {\it Phys.~Lett.} B212 (1988) 391.  \cr
\+ [2&9] &R.D.~Ball, S.~Forte and G.~Ridolfi, {\it Phys.~Lett.} B378 (1996)
255.  \cr
\+ [3&0] &SMC collaboration, {\it Phys.~Lett.} B412 (1997) 414. \cr  
\+ [3&1] &G.~Altarelli, R.D.~Ball, S.~Forte and G.~Ridolfi, {\it Nucl.~Phys.}
B496 (1997) 337.  \cr
\+ [3&2] &SMC collaboration, {\it Phys.~Rev.} D58 (1998) 112001. \cr 
\+ [3&3] &G.M.~Shore, {\it in} Proceedings, Erice 1998 (to appear). \cr 
\+ [3&4] &G.M.~Shore and G.~Veneziano, {\it Nucl.~Phys.} B516 (1998) 333;  \cr
\+\ &{} &D.~De Florian, G.M.~Shore and G.~Veneziano, Proceedings of 1997 
Workshop  \cr
\+\ &{} &~~~~on Physics with Polarised Protons at HERA, hep-ph/9711358.\cr

\vfill\eject

\bye